\documentclass[onecolumn]{pasj00}
\SetRunningHead{T. Nakajima et al.}{New 60-cm Radio Survey Telescope}
\Received{2006/12/02}
\Accepted{2007/05/26}
\Published{2007/10/25}
\begin{document}
\title{New 60-cm Radio Survey Telescope with\\
the Sideband-Separating SIS Receiver for the 200 GHz Band}
\author{Taku \textsc{Nakajima}\altaffilmark{1}, Masahiro \textsc{Kaiden}\altaffilmark{1,2}, Jun \textsc{Korogi}\altaffilmark{1}, Kimihiro \textsc{Kimura}\altaffilmark{1}, Yoshinori \textsc{Yonekura}\altaffilmark{1},\\
Hideo \textsc{Ogawa}\altaffilmark{1}, Shingo \textsc{Nishiura}\altaffilmark{2}, Kazuhito \textsc{Dobashi}\altaffilmark{2}, Toshihiro \textsc{Handa}\altaffilmark{3}, Kotaro \textsc{Kohno}\altaffilmark{3},\\
Jun-Ichi \textsc{Morino}\altaffilmark{4}, Shin'ichiro \textsc{Asayama}\altaffilmark{4}, and Takashi \textsc{Noguchi}\altaffilmark{4}}
\altaffiltext{1}{Department of Physical Science, Graduate School of Science, Osaka Prefecture University,\\
1-1 Gakuen-cho, Naka-ku, Sakai, Osaka 599-8531}
\altaffiltext{2}{Department of Astronomy and Earth Sciences, Tokyo Gakugei University,\\
4-1-1 Nukuikita-machi, Koganei, Tokyo 184-8501}
\altaffiltext{3}{Institute of Astronomy, The University of Tokyo, 2-21-1 Osawa, Mitaka, Tokyo 181-0015}
\altaffiltext{4}{National Astronomical Observatory, 2-21-1 Osawa, Mitaka, Tokyo 181-8588}
\email{s\_tac@p.s.osakafu-u.ac.jp}
\maketitle

\begin{abstract}
We have upgraded the 60-cm radio survey telescope located in Nobeyama, Japan. We developed a new waveguide-type sideband-separating SIS mixer for the telescope, which enables the simultaneous detection of distinct molecular emission lines both in the upper and lower sidebands. Over the RF frequency range of 205--240 GHz, the single-sideband receiver noise temperatures of the new mixer are 40--100 K for the 4.0--8.0 GHz IF frequency band. The image rejection ratios are greater than 10 dB over the same range. For the dual IF signals obtained by the receiver, we have developed two sets of acousto-optical spectrometers and a telescope control system. Using the new telescope system, we successfully detected the $^{12}$CO ($J=2-1$) and $^{13}$CO ($J=2-1$) emission lines simultaneously toward Orion KL in 2005 March. Using the waveguide-type sideband-separating SIS mixer for the 200 GHz band, we have initiated the first simultaneous $^{12}$CO ($J=2-1$) and $^{13}$CO ($J=2-1$) survey of the galactic plane as well as large-scale mapping observations of nearby molecular clouds. 
\end{abstract}

\section{Introduction}

\begin{figure}[ht]
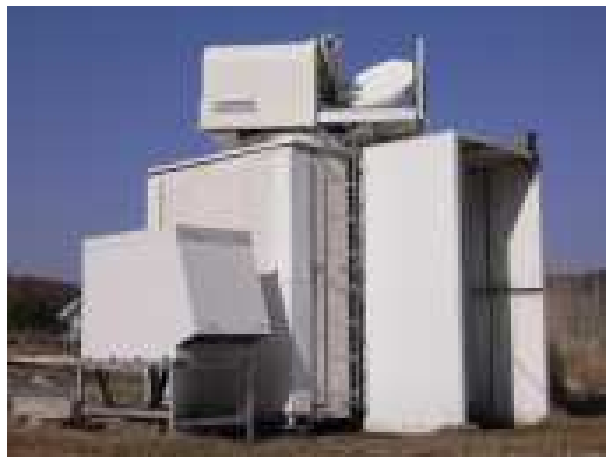

  \begin{center}
  \FigureFile(80mm,60mm){figure1.eps}
  \end{center}
 \caption{An overview of the 60-cm survey telescope (VST-1) in Nobeyama.}\label{fig1}
\end{figure}

The carbon monoxide (CO) emission lines are the principal probe used to investigate the distribution, structure, and kinematics of interstellar molecular gas in the Galaxy. In order to investigate the global structure of extended molecular clouds, a small-aperture telescope has an advantage, because its large beam makes it possible to obtain fully-sampled maps of large regions quickly. Therefore, there are several small radio telescopes with which to study molecular clouds extensively. For example, the 1.2-m telescopes with a beam size of 8.$\!^{\prime}$7 at the Harvard-Smithsonian Center for Astrophysics (CfA) and Cerro Tololo Inter-American Observatory in Chile have been used to survey the Milky Way in the $^{12}$CO ($J=1-0$) emission line (e.g., Dame \& Thaddeus\ 1985; Dame et al.\ 1987;\ 2001). The Very Small Telescopes with a diameter of 60 cm in Nobeyama (Very Small Telescope 1: VST-1) and in Chile (Very Small Telescope 2: VST-2) of the Institute of Astronomy at the University of Tokyo, having the same beam size as the CfA 1.2-m telescopes but tuned to the $^{12}$CO ($J=2-1$) emission line, were also employed for a survey of the galactic plane, nearby molecular clouds, and the Large Magellanic Cloud (e.g., Sakamoto et al.\ 1994;\ 1995; Sorai et al.\ 2001). Large-scale observations with the 4-m telescope in Nagoya and the NANTEN telescope in Chile of Nagoya University having a moderate beam size of 2.$\!^{\prime}$7 have revealed global structures of nearby molecular clouds as well as dense cores inside them in $^{12}$CO, $^{13}$CO, and C$^{18}$O ($J=1-0$) (e.g., Fukui \& Yonekura\ 1998; Onishi et al.\ 2004). Mapping observations with higher rotational transitions of CO and its isotopes were carried out by the Mt. Fuji 1.2-m telescope in the $J=3-2$ transition of $^{12}$CO (HPBW=3.$\!^{\prime}$1; e.g., Ikeda et al.\ 1999) as well as by the K\"{o}lner Observatorium f\"{u}r Submillimeter Astronomie (KOSMA) 3-m telescope in the $J=2-1$ and $J=3-2$ transitions of $^{12}$CO, $^{13}$CO, and C$^{18}$O (Kramer et al.\ 1996; Beuther et al.\ 2000). The beam size of the latter telescope is 2.$\!^{\prime}$1 at 200 GHz and 1.$\!^{\prime}$2 at 345 GHz.

For spectral line observations, it is important to operate the heterodyne mixer receiver in the single-sideband (SSB) mode rejecting the power from the image sideband, because an atmospheric noise coming from the image sideband increases the system noise temperature significantly. In the case of mixer receivers operated in the double-sideband (DSB) mode, it is difficult to perform an accurate intensity calibration using the ``chopper wheel'' method (Ulich \& Haas\ 1976; Kutner \& Ulich\ 1981), because it is difficult to achieve precise measurement of the image rejection ratio (IRR). Most telescopes, therefore, adopt one of the following techniques to operate the mixer receivers in the SSB mode: (1) a mechanically tunable interferometer as an image rejection SSB filter, (2) a mixer block equipped with backshort tuners, and (3) a sideband-separating (2SB) method by quasi-optics using a wire grid, two focusing mirrors, and two mixer horns. The VST-1 was equipped with an quasi-optical SSB filter before the development introduced in the present paper. Each technique requires quasi-optics and/or tuning mechanics, which makes the gain of the receivers unstable, and the loss in the optics increases the total noise temperature of the receiver systems. Moreover, each SSB method except for the 2SB can access only one sideband simultaneously, i.e., the upper sideband (USB) or the lower sideband (LSB), which is a well-known feature of the heterodyne mixer receivers in the SSB mode. 

\begin{figure}[hb!]
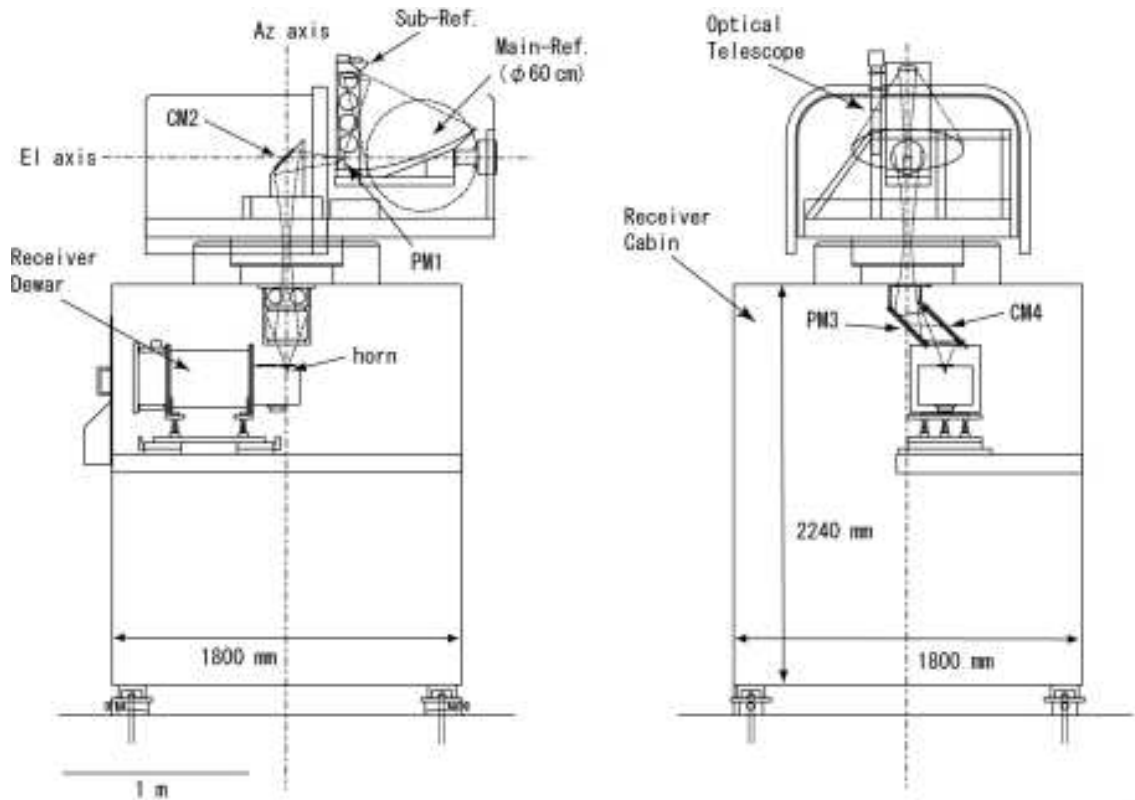

  \begin{center}
  \FigureFile(150mm,113mm){figure2.eps}
  \end{center}
  \caption{Mechanical structure and optics of the VST-1. A main reflector of 60 cm in diameter is set up on the rooftop of the receiver cabin, which is 1,800 mm in width and depth and 2,240 mm in height. The receiver dewar is fixed on a table in the cabin and is located at the Coud\'{e} focus. The two mirrors PM1 and PM3 are plain, and the two mirrors CM2 and CM4 are ellipsoid.}\label{fig2}
\end{figure}

In order to improve such disadvantages of the heterodyne mixer receivers, we have developed a waveguide-type 2SB Superconductor-Insulator-Superconductor (SIS) mixer receiver for the 200 GHz band and installed in VST-1. The waveguide-type 2SB receiver is free from the quasi-optical SSB filter and has ability to observe signals in the two sidebands independently and simultaneously. This important feature of the 2SB receiver should enable us to observe two or more spectral lines that are separated by more than 10 GHz at the same time, and the performance of the observations will be greatly improved. In addition, it is also noteworthy that through the simultaneous detection, we can better determine the line intensity ratios of two molecular lines, e.g., $T$($^{12}$CO)/$T$($^{13}$CO), without being contaminated by errors in pointing.

For the dual intermediate frequency (IF) signals obtained by the 2SB receiver, we have developed two sets of acousto-optical spectrometers (AOSs) and upgraded the telescope control system. Using the renewed VST-1 telescope, we simultaneously detected the $^{12}$CO ($J=2-1$ : 230.538 GHz) line in USB and the $^{13}$CO ($J=2-1$ : 220.398 GHz) line in LSB toward Orion KL in March 2005. To date, simultaneous detection of molecular emission lines with a waveguide-type 2SB receiver has been performed by a few groups toward limited number of directions in the sky (e.g., Asayama et al.\ 2003; Lauria et al.\ 2006), suggesting that the 2SB receivers are more efficient in comparison with other SSB receivers. Using the upgraded VST-1 telescope, we have initiated a survey of molecular clouds with the $^{12}$CO ($J=2-1$) and $^{13}$CO ($J=2-1$) emission lines. These are the first large-scale observations with the waveguide type 2SB SIS mixer for the 200 GHz band.

In the present paper, we describe the instrumentation of the upgraded VST-1 60-cm radio survey telescope and demonstrate its performance. The results of the test observation using the 2SB SIS mixer are also shown. The designs of the telescope and related instruments are described in \S 2. Measurements of the pointing accuracy, beam size, and beam efficiency are given in \S 3. In \S 4, we present the results of the test observations toward giant molecular clouds in the Orion-Monoceros region.

\begin{table}[ht!]
\begin{center}
\caption{VST-1 specifications}
\begin{tabular}{cc}
  \hline\hline
  {} & {}\\
  Parameter & Value \\
  {} & {}\\
  \hline
  {} & {}\\
  \underline{Main reflector}\\
  {} & {}\\
  Optics & offset paraboloid\\
  Aperture diameter ($d$)& 60 cm\\
  Focal length ($f$)& 60 cm\\
  Focal ratio ($f/D$)\footnotemark[$*$] & 0.4 (prime focus)\\
  Surface accuracy & 16 $\mu$m rms\\
  {} & {}\\
  \underline{Receiver}\\
  {} & {}\\
  Mixer type & sideband-separating SIS\\
  Tuning range & 208--230 GHz (LSB)\\
  {} & 220--242 GHz (USB)\\
  IF frequency & 4.0--8.0 GHz (1st IF)\\
  Noise temperature & ${\lesssim}$100 K (SSB)\\
  {} & {}\\
  \underline{Spectrometers}\\
  {} & {}\\
  Type & AOS\\
  Bandwidth & 250 MHz $\times$ 2\\
@Channel number & 2,048 channels $\times$ 2\\
  Resolution & 230 kHz\\
  {} & {}\\
  \hline
  \multicolumn{2}{@{}l@{}}{\hbox to 0pt{\parbox{85mm}{\footnotesize
      \par\noindent
      \footnotemark[$*$] $D$ is the outer diameter from the geometrically symmetric axis and differs from aperture diameter ($d$) because the VST-1 has offset Cassegrain optics (see \S 2.1). 
    }\hss}}
\end{tabular}
\end{center}
\end{table}

\section{Instruments}

The VST-1 60-cm radio survey telescope (Figure 1) was constructed by the Institute of Astronomy of the University of Tokyo and the Nobeyama Radio Observatory of the National Astronomical Observatory of Japan during 1988--1989. It is located at Nobeyama Radio Observatory in Japan (Hayashi et al.\ 1990). The primary purpose of this telescope was a large-scale survey in the $^{12}$CO ($J=2-1$) emission line to determine the temperature and density distribution of molecular gas in the Galaxy by comparing the $^{12}$CO ($J=2-1$) data with the $^{12}$CO ($J=1-0$) data obtained with the CfA 1.2-m survey (Dame et al.\ 1987). The beam size of the VST-1 was designed to be the same as that of the CfA 1.2-m telescope, so that one can directly compare the data from the two telescopes.

Two complementary observational projects were carried out with VST-1. One is the $^{12}$CO ($J=2-1$) survey of the northern galactic plane (e.g., Sakamoto et al.\ 1995,\ 1997; Oka et al.\ 1996,\ 1998) and the other is the $^{12}$CO ($J=2-1$) survey of nearby clouds (e.g., Sakamoto et al.\ 1994; Seta et al.\ 1998). The former was performed in order to understand the galactic scale variation of the physical conditions of molecular gas, whereas the latter was to understand the cloud-scale variation and to provide the templates necessary for an appropriate interpretation of the results of the galactic plane survey through a population synthesis of molecular gas. 

However, the receiver system of the VST-1 was not completely sufficient for operation, due to the quasi-optical SSB filter. For example, the SSB filter is too sensitive for its optical alignment. Poor positional reproducibility often increases the loss in the optics after maintenance of the receiver. In the worst case, the total system noise temperature is $\sim$10 times higher than the noise temperature of the mixer unit. In addition, only one sideband could be observed simultaneously, because the other sideband was terminated for the SSB mode.

To solve these problems, we have started a renovation of the telescope, including the replacement of the receiver system, in 2003. The installed 2SB receiver enable us to observe the $^{12}$CO ($J=2-1$) and $^{13}$CO ($J=2-1$) emission lines simultaneously. We also upgraded the optics, IF systems, spectrometers, and operation software. In the following subsections, we briefly describe the components of the new system. The main features of the telescope are summarized in Table 1. 

\begin{table*}[ht!]
\begin{center}
\caption{Parameters of the new optics.}
\begin{tabular}{llcc}
  \hline\hline
  {} & {} & {} & {}\\
  \multicolumn{2}{c}{Parameter} & \multicolumn{2}{c}{[mm]} \\
  {} & {} & 220 GHz & 230 GHz\\
  {} & {} & {} & {}\\
  \hline
  {} & {} & {} & {}\\
  Distance between the subreflector (SR) and the ellipsoidal mirror (CM2) & $L_{\mathrm{1}}$ & 765.77 & 765.77\\
  Distance between CM2 and the ellipsoidal mirror (CM4) & $L_{\mathrm{2}}$ & 1115.47 & 1115.47\\
  Distance between CM4 and the horn & $L_{\mathrm{3}}$ & 230.00 & 230.00\\
  Focal length of CM2 & $f_{\mathrm{1}}$ & 197.45 & 197.45\\
  Focal length of CM4 & $f_{\mathrm{2}}$ & 180.99 & 180.99\\
  {} & {} & {} & {}\\
  Gaussian beam radius at SR & $\omega_{\mathrm{SR}}$ & 38.27 & 38.27\\
  Gaussian beam radius at CM2 & $\omega_{\mathrm{1}}$ & 26.13 & 26.01\\
  Gaussian beam radius at CM4 & $\omega_{\mathrm{2}}$ & 35.98 & 35.06\\
  Gaussian beam radius at the horn & $\omega_{\mathrm{horn}}$ & 3.600 & 3.600\\
  {} & {} & {} & {}\\
  Beam waist between SR and CM2 & $\omega_{\mathrm{0}}$-1 & 5.23 & 5.01\\
  Beam waist between CM2 and CM4 & $\omega_{\mathrm{0}}$-2 & 8.07 & 7.85\\
  Beam waist between CM4 and the horn & $\omega_{\mathrm{3}}$-3 & 2.95 & 2.91\\
  {} & {} & {} & {}\\
  Curvature of SR & $R_{\mathrm{SR}}$ & 465.76 & 465.76\\
  Curvature of CM2 toward SR & $R_{\mathrm{11}}$ & 321.57 & 319.82\\
  Curvature of CM2 toward CM4 & $R_{\mathrm{12}}$ & 511.54 & 516.05\\
  Curvature of CM4 toward CM2 & $R_{\mathrm{21}}$ & 687.39 & 680.49\\
  Curvature of CM4 toward the horn & $R_{\mathrm{22}}$ & 245.68 & 246.57\\
  Curvature of the horn & $R_{\mathrm{horn}}$ & 42.79 & 42.79\\
  {} & {} & {} & {}\\
  Aperture radius of the corrugated feed horn & $a$ & 5.63 & 5.63\\
  Length of the corrugated feed horn & $l$ & 42.42 & 42.42\\
  {} & {} & {} & {}\\
  \hline
\end{tabular}
\end{center}
\end{table*}

\subsection{Antenna optics}

\begin{figure}[h!]
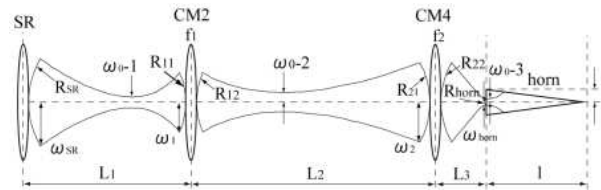

  \begin{center}
  \FigureFile(80mm,24mm){figure3.eps}
  \end{center}
 \caption{Definition of the optics parameters for a Gaussian beam.}\label{fig3}
\end{figure}

The mechanical structure and optics of the VST-1 after the upgrade are shown in Figure 2. The antenna mount and the optics from the primary mirror to the ellipsoidal mirror (CM2) are the same as the original. The offset Cassegrain-Coud\'{e} optics without blockage by a subreflector and its stay provided us with very high aperture efficiency and sufficiently flat baselines, which are important for observing broad and weak emission lines seen in the galactic plane. The main reflector is an offset paraboloid with a diameter of $d$=60 cm and a focal length of $f$=60 cm. The shape of the main reflector is part of a paraboloid ring, whose inner and outer radii from the symmetric axis are 15 cm and 75 cm, respectively. The curvature of the mirror has the same shape of an axisymmetric paraboloid with $D$=150 cm and $f$=60 cm, which gives $f/D$=0.4 for the prime focus. The surface accuracy of the aluminum main reflector is measured to be 16 $\mu$m rms ($\lambda$/81 at 1.3 mm) using a parabolic template and a feeler gauge. The synthetic focal length ($F$) of the main reflector and subreflector system is 300 cm. The effective $F/d$ ratio is 5 at the Cassegrain focus.

The beam of the Cassegrain focus is led to the Coud\'{e} focus with two ellipsoidal mirrors and two plane mirrors. In order to adjust the focal ratio between the antenna optics and the feed horn of the new receiver (see \S 2.2), the relay optics in the receiver cabin were re-designed. Assuming the edge taper level of the subreflector to be -12 dB and the frequency independent matching between the subreflector and the feed horn, we designed the new optics based on a Gaussian beam propagation (Chu\ 1983). The parameters of the new optics are summarized in Table 2, and their definitions are shown in Figure 3. 

A sliding cover protects the optics against rain and snow. By opening the sliding cover, we can begin observation immediately after the weather becomes better.

\subsection{Receiver system}

\begin{figure}[ht]
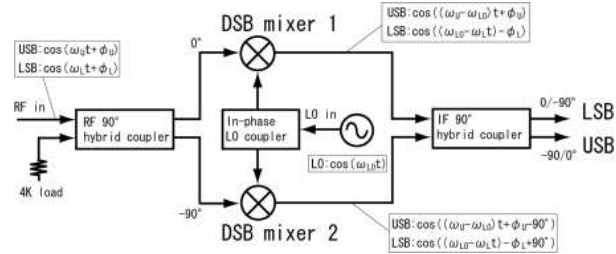

  \begin{center}
  \FigureFile(80mm,33.5mm){figure4.eps}
  \end{center}
 \caption{Block diagram of the 2SB mixer.}\label{fig4}
\end{figure}

\begin{figure}[ht]
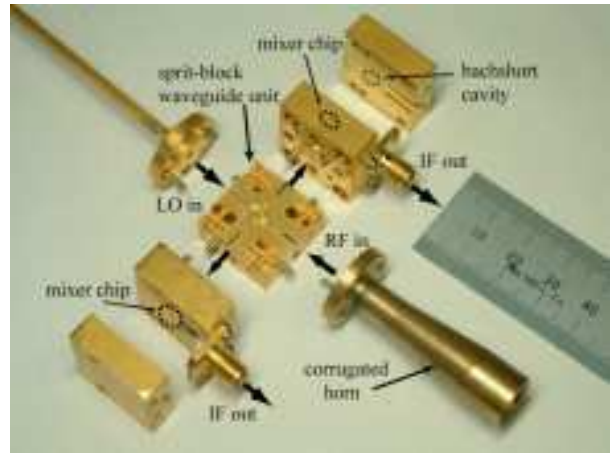

  \begin{center}
  \FigureFile(80mm,40mm){figure5.eps}
  \end{center}
 \caption{Photograph of the assembled 2SB mixer.}\label{fig5}
\end{figure}

\begin{figure}[ht]
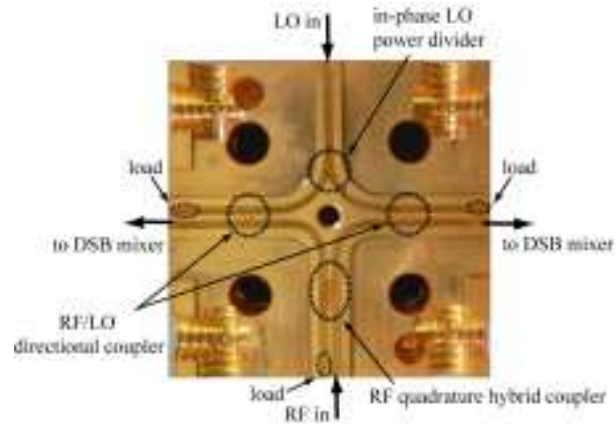

  \begin{center}
  \FigureFile(80mm,60mm){figure6.eps}
  \end{center}
 \caption{Photograph of the split-block waveguide unit.}\label{fig6}
\end{figure}

\begin{figure}[ht]
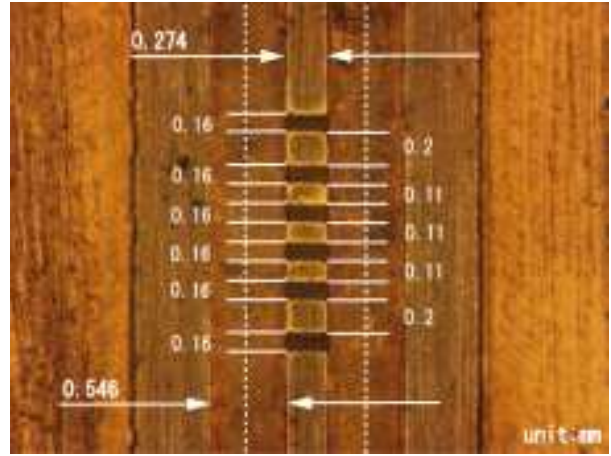

  \begin{center}
  \FigureFile(80mm,60mm){figure7.eps}
  \end{center}
 \caption{Microscopic image of the RF quadrature hybrid coupler. Separation of the input microwave signal into two outputs with equal intensity and a 90-degree phase difference is achieved using a waveguide quadrature hybrid with a six branch line coupler.}\label{fig7}
\end{figure}

The block diagram of the 2SB mixer is shown in Figure 4. We have chosen a modular approach for the 2SB mixer, which was originally proposed by Claude et al.\ (2000). The 2SB mixer consists of a split-block waveguide unit and two DSB mixers. The detailed structure of the assembled 2SB mixer is shown in Figure 5. The radio frequency (RF) signal is fed into the waveguide unit located in the middle of the figure from the lower-right side using a detachable corrugated feed horn. The local oscillator (LO) signal is introduced from the opposite side. An RF quadrature hybrid coupler, an in-phase LO power divider, two RF/LO directional couplers, and 4 K cold image terminations constitute one split-block waveguide unit (Figure 6), and each DSB mixer is separated from the split-block waveguide unit. This enables the DSB mixers to be evaluated separately from the waveguide unit prior to the integration. The RF/LO coupler is the same as the scaled model described by Asayama et al.\ (2004). A close-up picture of the RF waveguide quadrature hybrid coupler is shown in Figure 7. The IF signals from the two DSB mixers are combined in a commercial quadrature hybrid constructed by Anaren Microwave Inc. The direct-current (DC) bias for each of the DSB mixers is supplied independently using coaxial bias tees via the IF hybrid coupler.

The DSB SIS mixers adopted here were developed at the Nobeyama Radio Observatory. The SIS mixer design is based on the proposed bow-tie waveguide probe with a parallel-connected twin-junction (PCTJ) (Noguchi et al.\ 1995). The main difference between the present mixer and conventional mixers is the use of the half-reduced height waveguide for waveguide-to-stripline transition of the SIS mixer which is designed using the lumped-gap-source port provided by HFSS$^{\rm TM}$ (Asayama et al.\ 2003). The measured DSB receiver noise temperatures of the SIS mixers with 4.0--8.0 GHz IF are less than 50 K over the RF frequency range of 213--234 GHz. 

Before installing the receiver in the telescope, we evaluated its performance. The noise temperature of the 2SB SIS receiver was measured by a standard Y-factor method using hot (300 K) and cold (77 K) loads in the laboratory. The mixer was mounted on a 4 K cold stage of a dewar. The first-stage IF amplifier is a 4 K cooled High Electron Mobility Transistor (HEMT) at the 4.0--8.0 GHz band. The equivalent noise temperature and the gain of the HEMT amplifier associated with an isolator were approximately 10 K and +30 dB, respectively. The following-stage amplifiers work at room temperature. The SSB receiver noise temperature, including the noise contribution from the vacuum window, the feed horn, and the IF amplifier chain, are measured to be less than 100 K over the RF range of 208--242 GHz, and the minimum value of $\sim$35 K is achieved at around 225 GHz. The mean value and error are 60.2 $\pm$ 4.8 K (Figure 8a). The error is mainly due to the deviations of the radiation temperatures from the physical temperatures of the hot and cold loads. In the calculation, the radiation temperatures are assumed to be 300.0 $\pm$ 1.7 K and 80.0 $\pm$ 1.7 K for the hot and cold loads, respectively. We note that the RF frequency characteristic of Trx hardly change because the contribution of such deviations to the derived Trx is basically uniform for all measurement points. If we omit such deviations, the error in Trx is estimated to be $\pm$ 1.0 K. The IRRs were measured by the relative amplitudes of the IF responses in the USB and LSB when injecting a continuous-wave (CW) signal (Kerr et al.\ 2001). The measured IRRs were more than 10 dB over the same RF range with the mean value of 15.0 $\pm$ 2.2 dB (Figure 8b). The error is mainly due to the accuracy of the measuring instruments and fluctuations in the radiation temperatures of the hot and cold loads.

\begin{figure}[ht!]
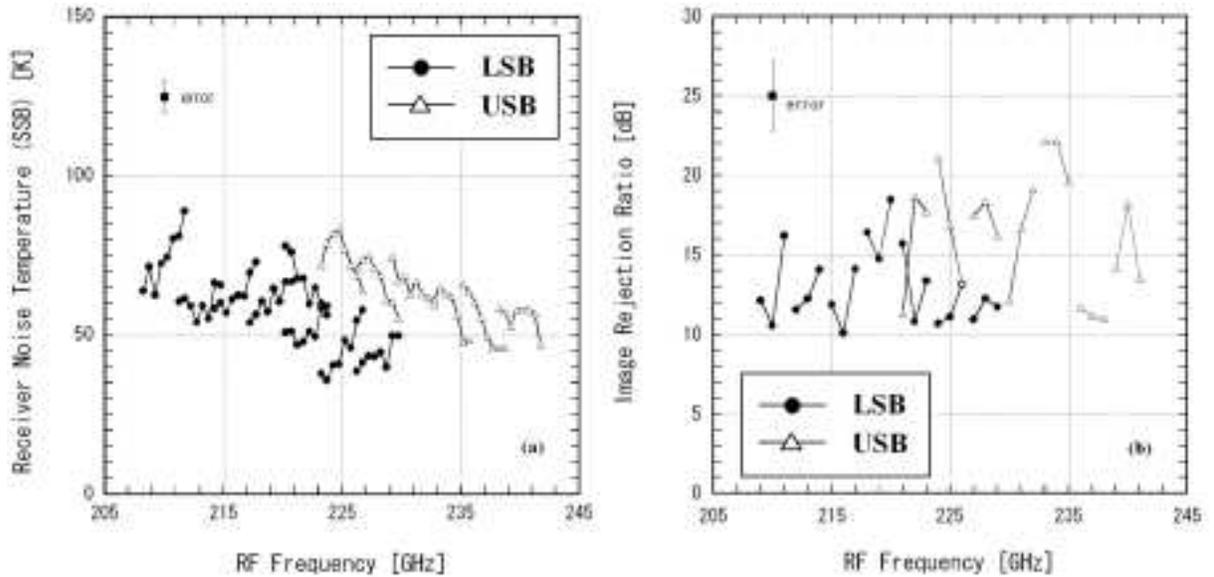

  \begin{center}
  \FigureFile(160mm,80mm){figure8.eps}
  \end{center}
 \caption{(a) The SSB receiver noise temperature and (b) the image rejection ratio in the LSB and USB are shown as functions of the RF frequency. The typical error is indicated in each figure (see text).}\label{fig8}
\end{figure}

\subsection{Intermediate frequency system}

For the dual IF signals obtained by the 2SB receiver, two independent IF chains, which connect the receiver and two sets of the back-end system (Figure 9), are required. The IF signals from the 2SB mixer are independently amplified by two cooled HEMT, one for each IF, attached to the 4-K stage in the dewar. Two isolators, one for each IF, are placed between the IF quadrature hybrid and the HEMT. The IF signals are amplified by +70 dB at room temperature and are converted from the first IF frequency range 4.0--8.0 GHz to the second IF frequency range 1.25--1.75 GHz and the third IF frequency range 250--500 MHz. The first LO frequency is 224.97 GHz and the first IF frequencies of emission lines are 5.568 GHz and 4.572 GHz for $^{12}$CO ($J=2-1$) and $^{13}$CO ($J=2-1$), respectively. A narrow band-pass filter with the bandwidth of 500 MHz and the stop-band attenuation of $>$ 20 dB is inserted just before the second mixer in each IF chain. The second and third LO signals are generated independently by four signal-generators (SG). The LO frequencies for two IF chains were designed to avoid any leaks of the $^{12}$CO and $^{13}$CO lines and the second LO signals into the other IF bands. The linearity of the IF system at room temperature as a whole and all main amplifiers is secured at least between 0 and $-$25 dB for the injection of 300 K radiation. Moreover, the 1-dB gain compression power is $\gtrsim$ 20 dB larger than the output power of all amplifiers and mixers for 300 K radiation. Because the injected radiation is at most 300 K, enough linearity is secured for the observation with this telescope.

Doppler frequency corrections are applied to the second LOs by adjusting the frequency of the SGs in the USB and LSB independently (see \S 2.5), because the frequency shifts are different between the two sidebands. The third IF signals are sent by coaxial cables to the console room through an underground tunnel. The lengths of the cables from the receiver cabin to the console room is approximately 70 m.

In the console room, the IF signals are amplified and injected to the spectrometers. At the input of the two spectrometers, each IF signal is divided into two signals. One is fed into the spectrometer, and the other is fed into the detector to monitor the total power of the IF signal. The amplitude of the IF signals passing into the spectrometers is approximately $+$15 dBm. 

\begin{figure}[ht]
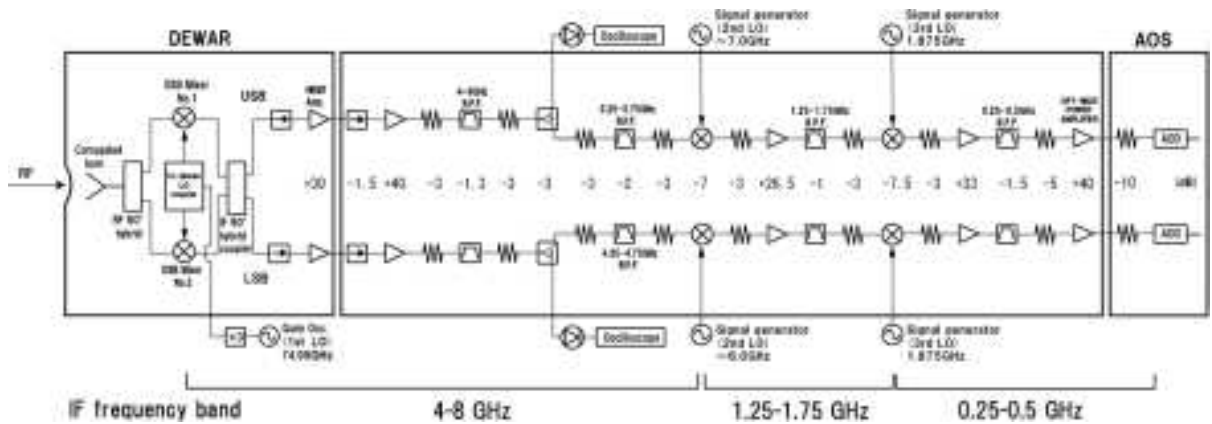

  \begin{center}
  \FigureFile(160mm,53.5mm){figure9.eps}
  \end{center}
 \caption{Block diagram between the receiver horn and AOSs.}\label{fig9}
\end{figure}

\subsection{Spectrometers}

\begin{figure}[ht]
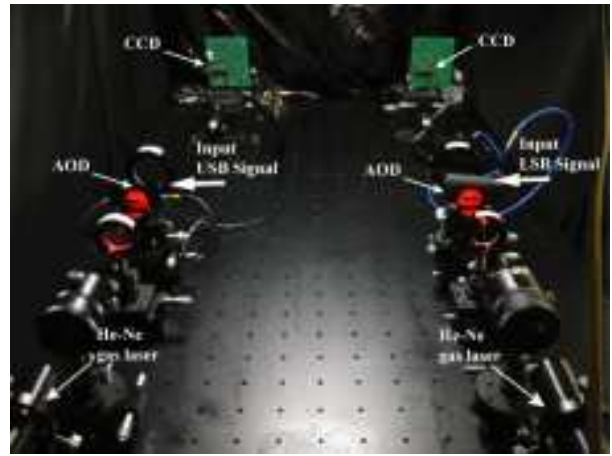

  \begin{center}
  \FigureFile(80mm,60mm){figure10.eps}
  \end{center}
 \caption{Photograph of two AOSs on the optical bench.}\label{fig10}
\end{figure}

In order to obtain spectral images in the USB and LSB simultaneously, we installed two sets of AOSs to the telescope. In an AOS, the IF signal from the receiver is converted into an ultrasonic wave by a piezo-electric transducer bonded to an acousto-optical material, and the wave travels through the material maintaining the spectral information of the RF signal. A monochromatic light beam from a laser tube is diffracted by the ultrasonic wave (i.e., Bragg reflection) and is focused to produce a spectral image on an imaging device (Lambert\ 1962).

Two AOSs are installed on an optical bench (90 cm ~ 180 cm) in a darkroom (Figure 10: Kaiden et al.\ 2006). The light-deflector element of the spectrometer, that is called the acoust-optical deflector (AOD), is a TeO$_{2}$ crystal, and a He-Ne gas laser tube at 632.8 nm is used as coherent light source. The first-order deflected light of each spectrometer is focused on a 2,048-channel charge coupled device (CCD) array. Each AOS has a bandwidth of 250 MHz. The frequency resolution is 230 kHz, corresponding to 0.3 km s$^{-1}$ at 230 GHz. The total velocity coverage is 335 km s$^{-1}$.

The Allan variance minimum time (Schieder et al.\ 1989) was found to be greater than 70 s. The AOSs are kept in an air-conditioned console room. We calibrated the channel-to-frequency relation of the AOSs several times a day and confirmed that the variation was less than one channel.

We developed a data-integration system for the AOSs. The analog signal outputs from the CCDs in both sidebands are converted into 12 bits of digital data with an A/D converter and sent to the integration-control PC. The spectral data of both sidebands are acquired and integrated at the same time on a single PC. The integration process runs on a Linux system. All integration procedures are written in the C language with some binary libraries provided by the maker of the digital I/O interface card. The control commands and the integrated data are transferred between the integration-control PC and the telescope-control PC (see \S 2.5) via TCP/IP.

\subsection{Control and data acquisition system}

\begin{figure}[ht]
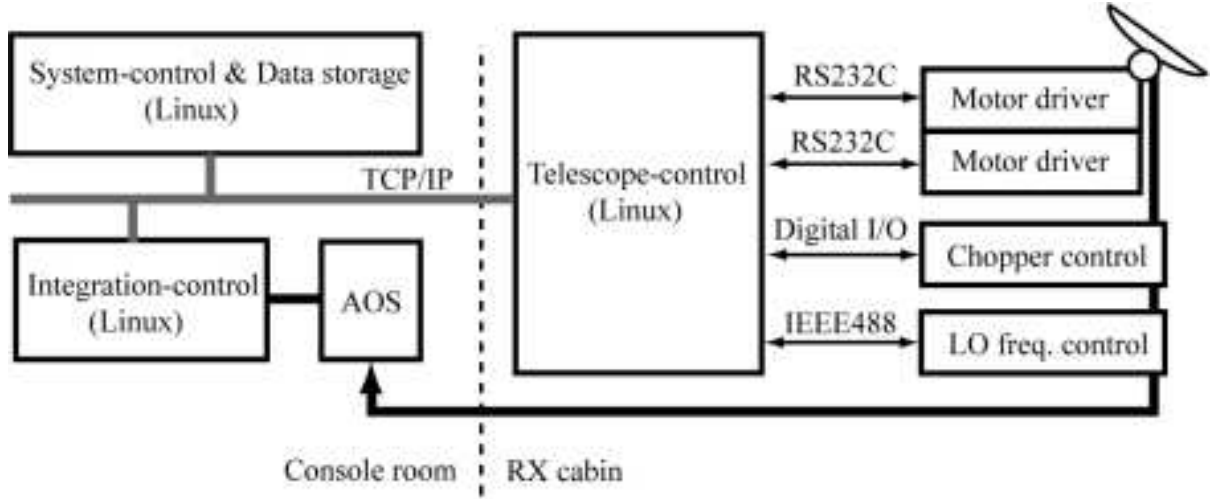

  \begin{center}
  \FigureFile(160mm,66mm){figure11.eps}
  \end{center}
 \caption{Schematic diagram of the VST-1 computer control system.}\label{fig11}
\end{figure}

We used a DOS-based telescope control system called {\it ASTROS}, before the installation of the 2SB SIS mixer receiver (Handa et al.\ 1996). The control system has been revised to a Linux-based system. The new system is called {\it UltraASTROS}, which is an acronym for ``UNIX/Linux-based Tools for Radio Astronomy - A Small Telescope Remote Observation System''. Because UltraASTROS is based on a multi-task operation system, it is composed of a number of tasks that run simultaneously. For example, the telescope monitor can be run as a different task of the telescope core operation task. The system is improved for easier accessibility through the internet, because TCP/IP services are supported on UNIX/Linux.

A hardware block diagram of the control system is illustrated in Figure 11. The drivers of the servomotors for elevation and azimuth tracking are connected via two RS-232C serial communication lines. Each axis of the telescope mount is driven directly by a servomotor made by NSK Co. Ltd. Each servomotor has a built-in angular encoder with a $\timeform{2".1}$ resolution. Astronomical calculations for tracking a star include the presession, the nutation by Sun and Moon, and the annual abberation due to perturbations of the Earth's orbit by Moon and four major planets. No ephemeris for the Moon or the planets was referenced. The refraction in millimeter waves due to the terrestrial atmosphere is corrected by a fomula with atmospheric temperature, pressure, and water vapor content used for the NRO 45-m telescope. For optical pointing observations the refraction in optical wave is corrected with another formula. It was originally developed for Subaru Telescope by Tanaka\ (1993) and we use his update (Tanaka\ 2006, private communication) of the formula replacing atmospheric refractive index given by Owens\ (1967). The Doppler tracking velocity is calculated using a formula given in a FORTRAN source code (Gordon\ 1976). The two SGs for the second LO are connected via an IEEE488 interface bus. The driver of the stepping motor for the chopper wheel is connected via digital I/O.

These instruments are controlled by the telescope-control PC in the receiver cabin. The telescope-control PC is connected to another PC called the system-control PC in the console room. Usually the system-control PC is the same machine as the data storage. The telescope-control PC and the system-control PC are connected by a 70 m long optical fiber. The two PCs communicate by TCP/IP through the optical fiber. After the correction of the atmospheric attenuation and the subtraction of the first-order baseline, the data are stored on the hard disk of the system-control PC for the reduction.

The clock for the telescope operation refers to the system clock of the telescope-control PC. It is adjusted several times a day by referring to the NTP server.

\subsection{Data reduction package}
In order to reduce the data observed with this telescope, we developed a software package called {\it UltraSTAR}, which is an acronym for ``UNIX/Linux based Tools for Radio Astronomy - a Streaming command set as Tools for Astronomical data Reduction''. UltraSTAR is a set of UNIX/Linux commands with which astronomers can access and process observed data.

With UltraSTAR, a user can access both spectral line data and three-dimensional cube data. These data are stored in the original formats, although some commands can convert the observed data from/to in popular formats such as FITS.

The main user interface of UltraSTAR is character based, because we believe data reduction procedures can be performed more efficiently through the keyboard. It is because the users apply the same parameters to different observational data, or different parameters to the same data in most cases. The parameters should be assigned with sufficient accuracy, which is not easily obtained through a graphical user interface. Only commands that require parameters with some allowance can be used through a graphical user interface (GUI).

In UltraSTAR, the data are stored in a file and only a few commands can access the data directly. Most of the commands are designed as filters for observational data, which are similar to several UNIX commands such as \texttt{sed} and \texttt{grep}. When the user wants to perform a complex process, this is possible through a sequence of filters connected by the command pipeline used in the UNIX system. For example, when a user want to remove spurious signals and apply baseline fitting to data stored in a file named ABC, the following command sequence should be given:

\texttt{ load ABC | spurious | baseline | save}\\
in UltraSTAR system.

Any wild-card characters in UNIX can be used in UltraASTROS file access commands. The data after any process will be stored under the same name in the default mode. The user must use the same name to access the latest data. Backup data are generated automatically to resume the process.

UltraSTAR was developed using open source software that can be obtained without any finantial charge. It can be run on UNIX or Linux systems.
The GUI interface was developed on LessTif. All source codes were written in C language. UltraSTAR itself is open source software and is available on the web at (\texttt{http://www.ioa.s.u-tokyo.ac.jp/VST/UltraSTAR/}).

\subsection{Optical pointing system}
The instrumental pointing errors are measured by observing approximately 150 stars using an optical telescope mounted on the subreflector stay. The optical telescope is a 5-cm refractor and the focal length is 40 cm. A monochromatic CCD camera for commercial use is installed on the prime focus. The measured pixel scales are 6.5 arcsec pixel$^{-1}$ for azimuth and 6.7 arcsec pixel$^{-1}$ for elevation. The output of the CCD camera is converted into a digital image by a video capture board on the system-control PC. The offset between the center of gravity of the stellar image and a reference position in the image is estimated using software that we developed. The estimated offset is automatically recorded with observational logging data by an optical pointing software running on the telescope-control PC.

\section{Performance}

\begin{figure}[ht]
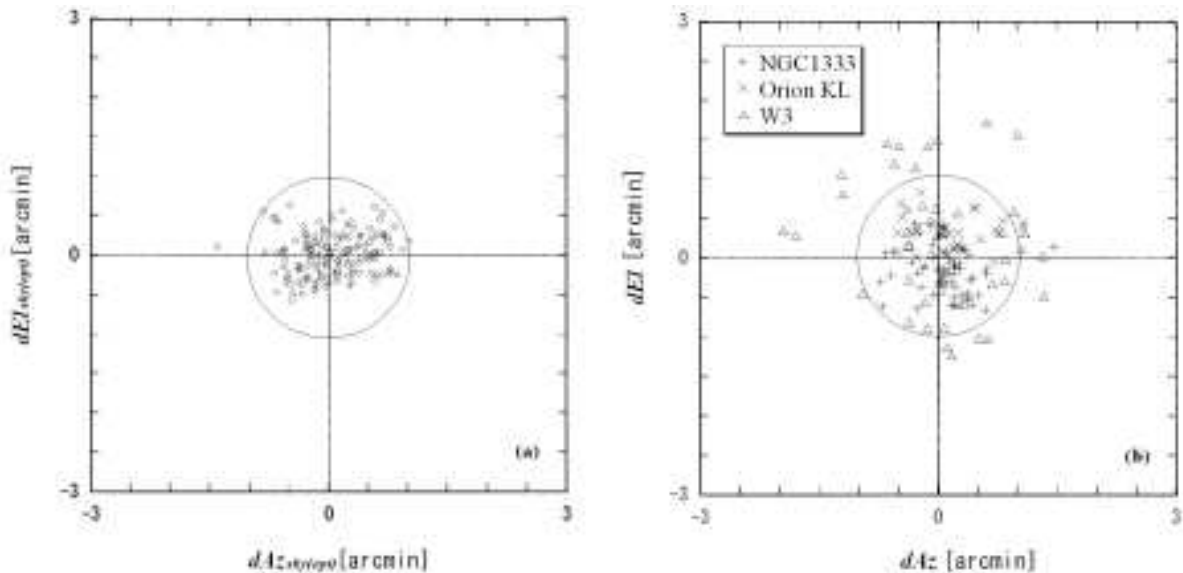

  \begin{center}
  \FigureFile(160mm,80mm){figure12.eps}
  \end{center}
 \caption{Scattering plots of (a) the optical pointing residuals and (b) the radio pointing residuals in azimuth and elevation. A circle with a radius of 1$^{\prime}$ is shown for comparison.}\label{fig12}
\end{figure}

\subsection{Pointing accuracy}

The telescope pointing is corrected using fixed parameters derived from pointing observations. We determined the pointing correction parameter using a procedure performed in two steps. The first step is performed by the observation of bright stars nominated from a standard position catalog of stars corrected with the proper motion of each star via an optical telescope. The second step is performed by the observations of radio sources with the receiver. No real-time correction due to the seeing is applied because of the large beam size.\\

\subsubsection{Optical pointing calibration}
The pointing error due to the antenna mount is measured by optical pointing. The pointing error should be removed using the following equations with six parameters (the flexure term is neglected for our system):
\begin{eqnarray}
dAz_{\rm sky(opt)} &=& dAz_{\rm enc(opt)} \cdot \cos (El) \nonumber \\
        &=& A_1 \sin (El) + A_2 + A_3 \cos (El) + B_1 \sin (Az) \sin (El) - B_2
\cos (Az) \sin (El)
\end{eqnarray}
and
\begin{eqnarray}
dEl_{\rm sky(opt)} &=& dEl_{\rm enc(opt)} \nonumber \\
&=& B_1 \cos (Az) + B_2 \sin (Az) + B_3,
\end{eqnarray}
where $dAz_{\rm sky(opt)}$ and $dEl_{\rm sky(opt)}$ are deviations along the azimuth $Az$ and elevation $El$ directions on the focal plane, and $dAz_{\rm enc(opt)}$ and $dEl_{\rm enc(opt)}$ are the same but measured at rotation axis, which is equivalent to the value at encoder. The parameter meaning is as follows. $A_{1}$ originates in the orthogonality of the azimuth and elevation axes, $A_{2}$ is the constant offset between radio-axis and optical telescope, and $A_{3}$ is the constant offset of the azimuth encoder. $B_{1}$ and $B_{2}$ are the direction of the azimuth axis from the zenith, and $B_{3}$ is sum of the offset of the elevation encoder and the offset between radio-axis and optical telescope. 

We measured pointing errors by optical pointing several times at night in 2005 December and 2006 January. Approximately two hours were required to observe 150 stars as a measurement set. The residual after the fitting is 0.$\!^{\prime}$52, which corresponds to approximately one-twentieth of the HPBW (Figure 12a). The residual in the optical pointing may be caused by the tracking error which originates from the fact that the tracking is achieved by position control with the interval of one second, i.e., the telescope moves to the calculated position only once a second.\\

\subsubsection{Radio pointing calibration}
Although the imperfactness of the alt-azimuth mount can be calibrated by the optical pointing measurement, the radio receiver located at the Coud\'{e} focus can have a pointing error that cannot be found by optical pointing calibration. The additional pointing error is mainly due to offset of the receiver horn location on the focal plain. As the first stage we have calibrated it through the receiver responce depending on azimuth rotation with a 300 K absorber on a half of the sub-reflector. After this adjustment and the optical pointing calibration, the receiver is placed such that the feed horn is aligned with the optical axis of the telescope, and pointing calibration by radio measurement is required. With many radio telescopes, radio pointing is carried out by observing a number of point sources, such as maser sources and quasars. However, no appropriate sources can be detected with the VST-1 because of the large beam size. Therefore, we use compact but non-point-like sources such as compact molecular clouds. We perform five point observations in $^{12}$CO ($J=2-1$) emission toward the NGC1333, Orion KL, and W3 with 3.$\!^{\prime}$75 or 5.$\!^{\prime}$0 grid spacing. The source structure and telescope pointing error can be separated using different dependence on the azimuth and elevation direction. Three minutes are required to make one five-point observation. Then, the relative deviation of the center of gravity is calculated from each five-point observation. The pointing model should be expressed by the following equations with six parameters and the assumed position error of each calibration source:
\begin{eqnarray}
dAz &=& C_1 \sin (Az-El) + C_2 \cos (Az-El) + D_1 +E_1 \cos (El) - E_2 \sin (El) + \delta Az(source,t)
\end{eqnarray}
and
\begin{eqnarray}
dEl &=& C_1 \cos (Az-El) - C_2 \sin (Az-El) + D_2 + E_1 \sin (El) + E_2 \cos(El) + \delta El(source,t),
\end{eqnarray}
where $C_{1}$ and $C_{2}$ correct for the errors owing to the misalignment of the receiver horn to the radio telescope collimation axis, $D_{1}$ and $D_{2}$ correct for the error for any collimation offsets between the optical telescope and the radio beam, $E _{1}$ and $E_{2}$ correct for the offsets between the third mirror and the elevation rotation axis, $\delta Az$ is the azimuthal pointing error due to a positional error of the assumed source position, and $\delta El$ is the elevational pointing error due to a positional error of the assumed source position. The source structure is fixed on the equatorial coordinate. 

\begin{figure}[ht]
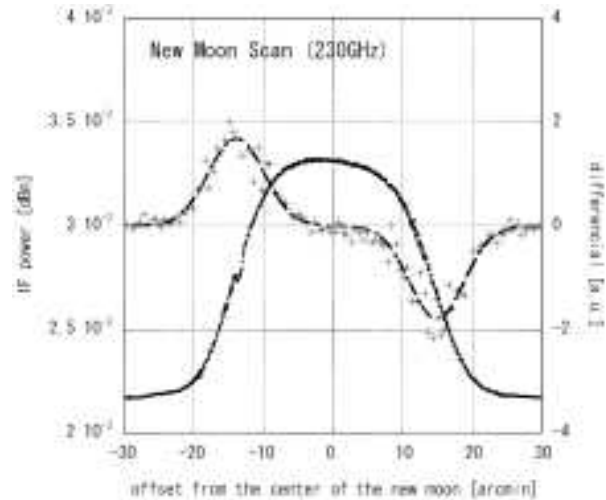

  \begin{center}
  \FigureFile(80mm,66mm){figure13.eps}
  \end{center}
 \caption{Total power of the IF output during the one-directional scan of the new moon (shown by dots). The plus symbols represent the results of the differential of the scanning data. The beam pattern is derived by the best fit to the differential assuming that the beam pattern has a Gaussian shape.}\label{fig13}
\end{figure}

Therefore, $\delta Az$ and $\delta El$ depend on the hour angle and declination of the source. After several measuring at different hour angles and declinations, we can get both the correct position of the sources and the telescope characteristic parameters. The detailed procedure will appear in a separated paper.

We measured the pointing errors by the radio pointing for three days in 2005 December. As many as 136 five-point observations were made at different azimuth and elevation. The pointing accuracy was measured to be 0.$\!^{\prime}$70 in rms, which is less than one-tenth of the HPBW (Figure 12b). The residual in the radio pointing may be caused by the fitting error due to the nonuniformity of the source position and the deviations of the beam pattern and the source brightness distribution from the assumed gaussian distribution as well as the causes described in the optical pointing.

\subsection{Beam size and efficiency}

We estimated the beam size and the efficiency of the telescope based on the observations of the new moon. We scanned the new moon in the azimuth and elevation directions, recording the total power of the third IF output (hereinafter referred to as scanning data). \\

\subsubsection{Beam size}
The beam pattern of the antenna has been obtained as the differential of the scanning data, if brightness distribution of a new moon is a uniform disk. Figure 13 illustrates the scanning data and the beam pattern in the azimuth direction. The half power beam width (HPBW) of the telescope is estimated to be 9.$\!^{\prime}$5 $\pm$ 0.$\!^{\prime}$8 at 230 GHz, assuming that the beam pattern has a Gaussian shape.\\

\subsubsection{Beam efficiency}

\begin{figure}[ht!]
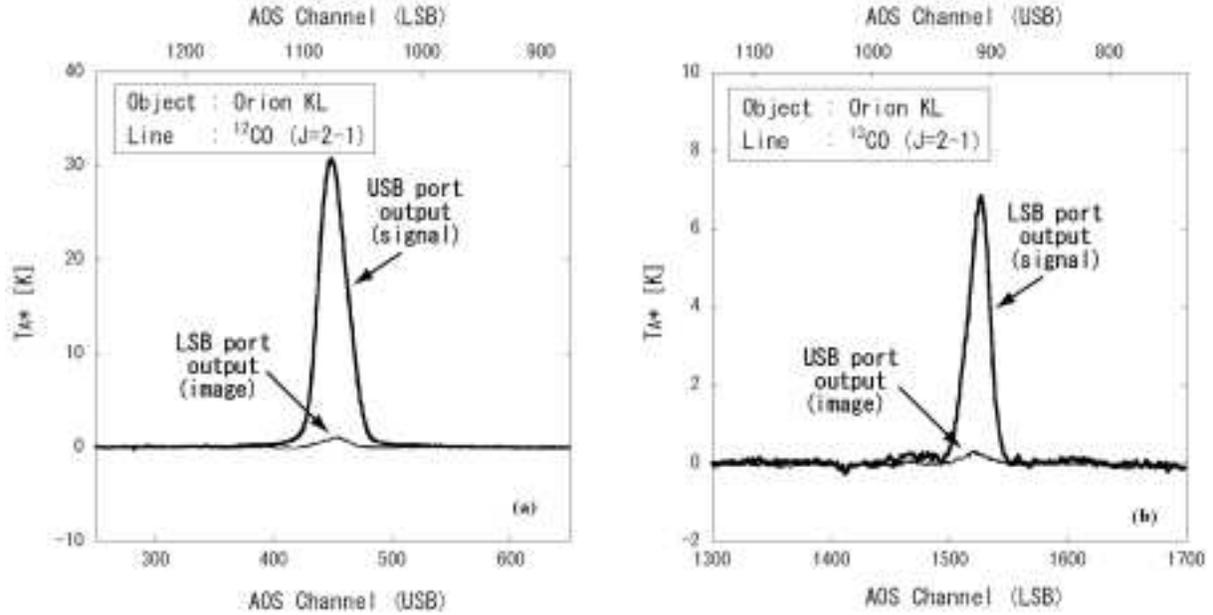

  \begin{center}
  \FigureFile(160mm,80mm){figure14.eps}
  \end{center}
 \caption{IF signal outputs of the USB and LSB ports obtained by observations toward Orion KL. (a) The signal of $^{12}$CO ($J=2-1$) in the USB port (bold line) leaks to the LSB port (thin line). (b) The signal of $^{13}$CO ($J=2-1$) in the LSB port (bold line) leaks to the USB port (thin line).}\label{fig14}
\end{figure}

The first side lobe level of the beam was designed to be less than $-$20 dB relative to the peak, and no significant side lobe was found in the derived beam pattern. We derived the main-beam efficiency ($\eta_{\mathrm{mb}}$) using the scanning data described above from the following equation, assuming that the brightness temperature distribution of the new moon is uniform:
\begin{eqnarray}
\eta_{\mathrm{mb}} = \frac{\int_{mainbeam}P d\Omega}{\int_{2\pi}P d\Omega} \sim \eta_{\mathrm{moon}} = \frac{\int_{moon}P d\Omega}{\int_{2\pi}P d\Omega} = \frac{T_{\mathrm{A}}^{*}}{T_{\mathrm{S}}},
\end{eqnarray}
where $T_{\mathrm{A}}^{*}$ is the measured antenna temperature of the new moon and $T_{\mathrm{s}}$ is the brightness temperature of the new moon, which is assumed to be 125 K at 230 GHz (Linsky 1973). Although the efficiency derived in this manner should be called as the Moon efficiency ($\eta_{\mathrm{moon}}$) and differs from $\eta_{\mathrm{mb}}$ in general, the difference between $\eta_{\mathrm{mb}}$ and $\eta_{\mathrm{moon}}$ is estimated to be $\lesssim$ 1 \% for the present telescope. Therefore we assume $\eta_{\mathrm{mb}}$ = $\eta_{\mathrm{moon}}$. The main-beam efficiency was estimated to be 97.4 $\pm$ 1.9 \% at 230 GHz.

\subsection{System noise temperature and image rejection ratio}

The performance of the receiver may change when it is installed in the telescope, because the environment of the receiver is different from that in the laboratory. We installed the receiver in the telescope and measured the performance of the receiver system. The typical SSB noise temperature of the system, including the atmosphere ($T_{\mathrm{sys}}$), ranged from 130 to 300 K at an elevation of 30 degrees during the observations. The noise temperature became approximately one-third of the previous receiver system. The IRRs were measured by observations of the $^{12}$CO ($J=2-1$) and $^{13}$CO ($J=2-1$) emission lines toward Orion KL in 2005 March. The corresponding IF signals at both ports are shown in Figure 14. The IRRs at the USB and LSB ports were 13.9 dB and 14.8 dB, respectively. These values are as good as those obtained with the previous system with the SSB filter. 

\subsection{Intensity calibration system}
The hourly gain variation of the receiving system was calibrated by chopping between the ambient temperature load and the sky. The intensities of raw data are presented in $T_{\mathrm{A}}^{*}$, the antenna temperature corrected for the atmospheric and antenna losses terminated at the ambient temperature (Ulich \& Haas 1976; Kutner \& Ulich 1981).

The absolute gain stability was monitored by observing Orion KL and M17, with which we can calibrate the daily change. We found that $T_{\mathrm{A}}^{*}$ gradually changes by $\pm$ 7.9 \% in the USB and $\pm$ 5.3 \% in the LSB over a time scale of several months. The difference in the fluctuation in Ta* for USB and LSB may be mainly caused by the pointing error, because the structure of the molecular clouds observed with the present telescope in $^{12}$CO is more compact than that of $^{13}$CO.

\section{Test observations}

\begin{figure}[ht]
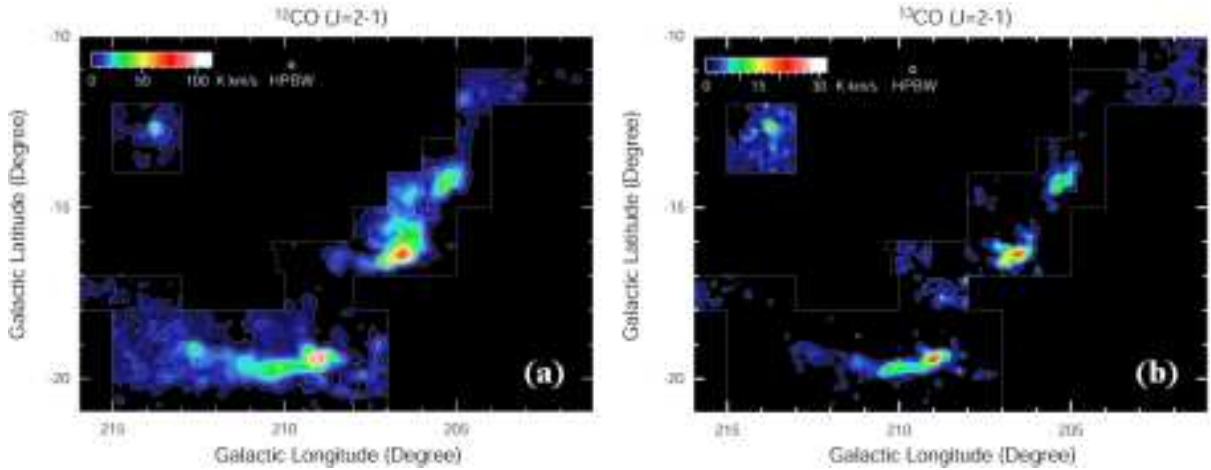

  \begin{center}
  \FigureFile(160mm,62mm){figure15.eps}
  \end{center}
 \caption{Integrated intensity maps of (a) $^{12}$CO ($J=2-1$) and (b) $^{13}$CO ($J=2-1$) toward the Orion region obtained by simultaneous observations. Integrated velocity ranges are from $-$10 km/s to 30 km/s for both of the maps. The lowest contour level and the contour intervals are (a) 3.0 K km/s and 6.0 K km/s, and (b) 2.2 K km/s and 4.4 K km/s, respectively. The observed regions are indicated by the boxes.}\label{fig15}
\end{figure}

After the first simultaneous detection of $^{12}$CO ($J=2-1$) and $^{13}$CO ($J=2-1$) with the new telescope system in 2005 March (see \S 3.3), we performed large-scale simultaneous mapping observations of the Orion region during 2006 February--April in order to check the total telescope system. The signals of $^{12}$CO ($J=2-1$) and $^{13}$CO ($J=2-1$) were led to the USB and the LSB, respectively, with the IF frequency of $\sim$5 GHz. The observation grid point is spaced by 7.$\!^{\prime}$5 along the galactic longitude and latitude. The observed region extends over approximately 55 square degrees. In total, approximately 3,500 spectra were obtained in each emission line. The achieved rms noise level was 0.5 K per channel after integrating the signal for 10 sec per position, which is consistent with the estimate from $T_{\mathrm{sys}}$. All line intensities in this section were presented in the antenna temperatures corrected for the atmospheric loss and the main-beam efficiency and are equal to the radiation temperature of a source filling the main beam.

The results are shown in integrated intensity maps in Figure 15, in which the integrated velocity range is between -10 and +30 km/s in $v_\mathrm{LSR}$. Emissions from Orion A ($l\sim$208--214 deg, $b\sim-$20--$-$18 deg), Orion B ($l\sim$204--208 deg, $b\sim-$17--$-$13 deg), and Mon R2 ($l\sim$213--215 deg, $b\sim-$14--$-$12 deg) are clearly observable in both maps. Other observations, both in the $^{12}$CO ($J=2-1$) and $^{13}$CO ($J=2-1$) emission lines of nearby molecular clouds, as well as of the northern galactic plane, are currently being performed, as will be reported in a subsequent publication.

\section{Conclusions}
We have completed the renovation of the VST-1 60-cm radio survey telescope, improving its sensitivity and observation efficiency. The drastic part of the renovation is the replacement of the existing receiver with the newly developed waveguide-type 2SB SIS receiver for the 200 GHz band. For the dual IF signals obtained by the 2SB receiver, we also designed and installed two sets of 2,048-channel acousto-optical spectrometers, both with bandwidths of 250 MHz. The existing DOS-based telescope control system was also replaced by a new Linux-based control system.

Using the renovated telescope, we successfully performed simultaneous CO multi-line observations and confirmed the performance of the new instruments. We confirmed the typical SSB noise temperature of the present system, including the atmosphere, became $\sim$ 1/3 of the previous system with an SSB filter. The observation time is therefore reduced to $\sim$ 1/9. In addition, the new receiver system enables the simultaneous detection of distinct molecular emission lines both in the USB and LSB, which makes another improvement in the observation efficiency at a factor of 1--2. In total, the observation efficiency became 10--20 times better than the previous system. The new system not only improves the observation efficiency, but also makes possible to observe both the $^{12}$CO ($J=2-1$) and $^{13}$CO ($J=2-1$) lines simultaneously with neither relative pointing error nor differences in atmospheric condition and performance of the telescope system. This may greatly improve the accuracy in deriving line ratios. We note that the installation of the present receiver system in other radio telescopes may also be meaningful to improve their performance. Simultaneous observations of $^{12}$CO ($J=2-1$) and $^{13}$CO ($J=2-1$) in the northern Galactic plane and nearby molecular clouds using this telescope have already begun.

\bigskip
The authors would like to thank Takafumi Kojima, Takahiro Yoda, Masaaki Hiramatsu, Seiichiro Naito, and Yasuhiro Abe for their contributions to this project. Tetsuo Hasegawa was an initial member of this project, and we would like to thank him for his help and advice. We are also grateful to Hiroyuki Iwashita, Toshikazu Takahashi, Akira Mori and the entire staff of the Nobeyama Radio Observatory for their useful discussions and support. This study was supported in part by Grants-in-Aid for Scientific Research C (17540214 and 18540232) from JSPS and by Grants-in-Aid for Scientific Research on Priority Areas (15071205, and 18026003) from MEXT.

\end{document}